\documentclass[aps,nofootinbib,superscriptaddress]{revtex4-2}
\usepackage[unicode=true,pdfusetitle,
 bookmarks=true,bookmarksnumbered=false,bookmarksopen=false,
 breaklinks=false,pdfborder={0 0 0},backref=false,colorlinks=true,citecolor=blue,urlcolor=blue 
]{hyperref}
\usepackage{xcolor}
\usepackage{graphicx}
\usepackage{orcidlink}
\usepackage{subfigure}

\begin{document}
\newcommand{\beq}{\begin{equation}}
\newcommand{\eeq}{\end{equation}}
\newcommand{\hlite}{\textcolor{violet}}

\bibliographystyle{apsrev}


\title{Diffusion Limited Aggregation in Ants Biting off Wall Paint}

\author{Suvrajyoti Chatterjee\,\orcidlink{0009-0004-5094-7013}}
\email{suvrajyoti.c@gmail.com}
\affiliation{SAP Labs India, 138, EPIP Zone Whitefield Rd, Whitefield, Bengaluru 560066, India.}
\author{Saba Firoze\,\orcidlink{0009-0009-3919-474X}}
\email{sabafiroze@gmail.com}
\affiliation{Sumadhura Aspire Aurum, Kannamangala Cross
Road, Bengaluru 560067, India.}
\author{Tabish Qureshi\,\orcidlink{0000-0002-8452-1078}}
\email{tqureshi@jmi.ac.in}
\affiliation{Centre for Theoretical Physics, Jamia Millia Islamia, New Delhi 110025, India.}


\begin{abstract}
Diffusion limited aggregation (DLA) is a well studied phenomenon in which
diffusing particles cumulatively aggregate on a starting fixed seed point,
forming a pattern which is fractal in structure. Here we report an interesting
DLA process arising from ants biting off wall paint. The structures 
arising from this long process are analyzed and found to have a fractional
dimension between 1 and 2.
The DLA formations observed in nature are mostly a result of various kinds
of inanimate {or micro-biological} particles diffusing and aggregating. What makes the present
observation interesting is that it involves live {insects} which may be guided
by sight, smell etc. A simple theoretical model of the process is simulated,
and the results agree well with the experimental observations.
\end{abstract}


\maketitle
\section{Introduction}

Pattern formation is a phenomenon common in nature, arising in widely
varying situations, from pattern formed in clouds, the spiral
formations of galaxies and hurricanes to the beautiful symmetries found
in snowflakes and silicon. Various processes contribute to the creation
of these natural patterns, typically involving an interplay between the
transport characteristics and the thermodynamic properties of the matter
and radiation involved.

Convection is generally the primary transport mechanism in both
terrestrial and astrophysical environments \cite{rayleigh}. 
However, in numerous natural scenarios, convection is not feasible. In
such instances, diffusion typically becomes the dominant transport
mechanism. Examples include the development of river networks, the
formation of frost on glass, and the arrangement of mineral veins
within geological structures. Likewise, convection is absent in many
laboratory processes, such as ion deposition, electrodeposition, and
other solidification methods.

\begin{figure}[h]
\centerline{\resizebox{10.0cm}{!}{\includegraphics{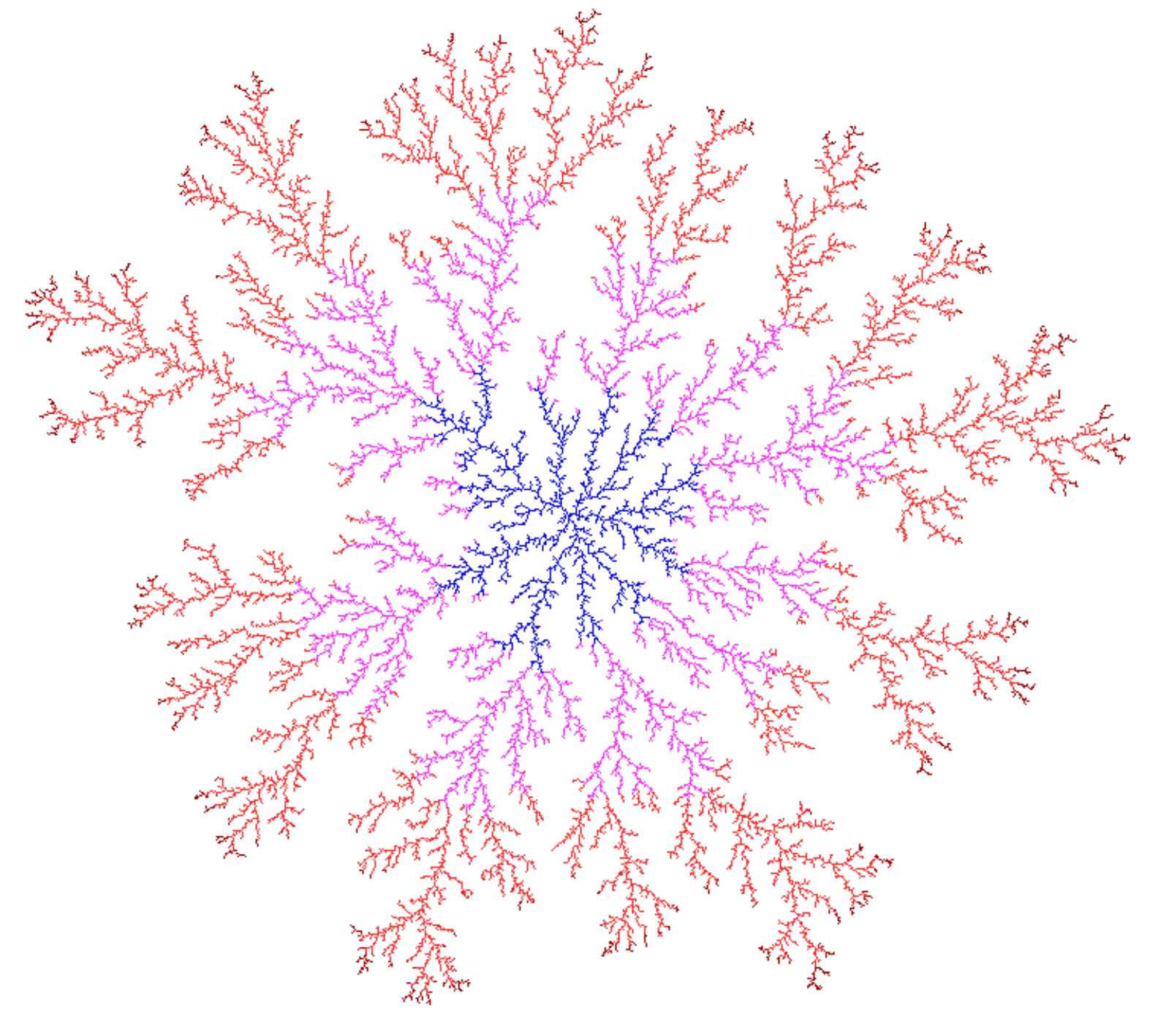}}}
\caption{A typical pattern formed in a computer simulation of DLA.
(Public-domain image from Wikipedia.)}
\label{dlatypical}
\end{figure}

The patterns that emerge in these systems exhibit certain common
characteristics, which can be described by several straightforward
models. The most notable of these is diffusion-limited aggregation (DLA)
\cite{dla}. The process involves having a single fixed point, and another
distant point moving via random diffusion. Eventually the second point 
comes close the first and sticks there. A third point then starts 
diffusing at a distance, and continues until it hits one of the two
stationary point, then it sticks there. A fourth diffusing particle
then starts, and the process thus goes on, involving a very large number
of particles. The structure thus formed is of a highly branched nature
(see Fig. \ref{dlatypical}), and posesses a fractional dimension, hence
referred to as a fractal \cite{mandelbrot}.
Initially proposed by Witten and Sander as a model for irreversible
colloidal aggregation \cite{witten}, it soon became apparent that DLA has
a broad range of applications. For example, DLA like patterns were observed
in electro-deposition of copper, in a geometry where there is a point
cathode in the center, and a ring-like anode surrounding it \cite{copper}.
In a completely unrelated setting, a procedure of oil extraction
from the ground, which involves pumping water into the region inside 
underground rocks containing oil, so that the water pushes out the oil
uniformly to aid extraction, led to the surprising result that water
formed progressively branching fingers, quite similar to the structures
seen in DLA. This phenomenon is known as Hele-Shaw fingering \cite{heleshaw}.
Hence DLA has come to be understood as a widely applicable model, and
has been a subject of sustained research
\cite{vicsek,ramaswamy,roder,somfai,sun,jungblut,halsey,halsey2,filtration}.

{DLA models have also been applied to some micro-biological systems like
aggregation of membrane molecules \cite{membrane}.
Microbial biofilms are
found in a variety of situations like in wastewater treatment systems. 
DLA models have been applied to describe the biofilm growth in such 
situations \cite{biofilm}.}

Here we report observation of DLA like pattern, formed as a result
of \emph{ants} biting off wall paint (see Fig. \ref{antsdla1}).
While the foraging
ants are easy to imagine as behaving like diffusing particles, the fact that
this phenonmenon does not involve particles following simple physical laws,
but involves {foraging} animals posessing sense of smell, sight etc, makes it
curiously interesting.

\section{Observations and analysis}

\subsection{Observations}

{Extensive ant activity was observed on a painted wall adjoining
a staircase leading to a terrace garden. Systematic observations
revealed that the ants were actively detaching minute paint fragments
and transporting them away. At any given time, numerous ants operated
concurrently within a localized region of the wall, with each individual
carrying a single paint fragment per foraging trip. Notably, the ants
were unable to initiate paint removal from intact, continuous paint
surfaces. Instead, they preferentially exploited pre-existing defects
or regions of minor paint deterioration. Beginning at these sites, the
ants progressively removed additional paint particles, thereby enlarging
the damaged areas. Over time, this collective activity generated distinct
patterns of varying sizes on the wall surface. Although individual removal
events were directly observable, the cumulative impact of the process
became evident only after several days to weeks due to its gradual nature.
}

\begin{figure}[t]
\subfigure[]{\includegraphics[width=5.7cm]{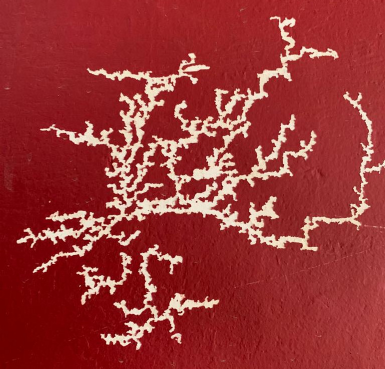}}
\subfigure[]{\includegraphics[width=5.7cm]{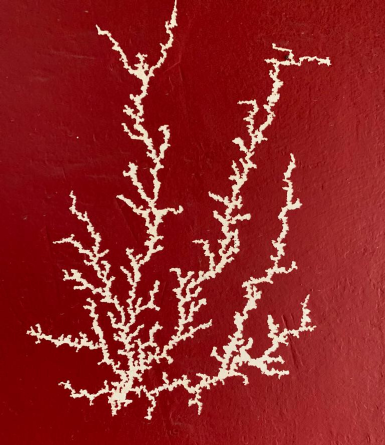}}
\subfigure[]{\includegraphics[width=5.7cm]{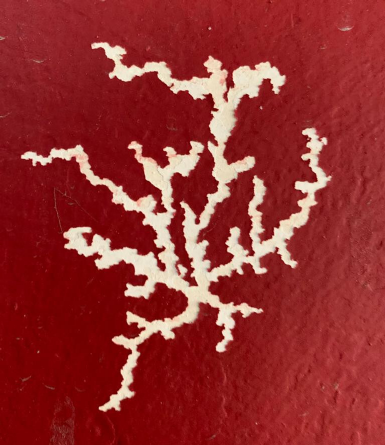}}
\caption{Patterns formed by ants biting off wall paint. Ants are able
to bite off paint only if there is already a chip in the paint. (a) This
is the most intricate of the three patterns. (b) Pattern
where a little chipped area was likely accessible to the ants just from
above and not from all sides. (c) Pattern where the paint is heavily
eaten away by the ants.}
\label{antsdla1}
\end{figure}

The observed patterns were typically 10-20 cm across. Some patterns were
constrained by the fact that ants could access an area only from one side,
and not from all sides. {The patterns were photographed to generate
images 800-1000 pixels in length and breadth.}



\subsection{Analysis}

{Fractional dimension is a concept used to describe objects
whose geometric complexity lies between two whole-number dimensions. In
ordinary geometry, a line has dimension 1, a surface has dimension 2,
and a solid has dimension 3. However, irregular and fractal structures,
such as coastlines, clouds, and branching patterns, may occupy space in
a way that cannot be represented by an integer dimension. The concept
of dimension can be generalized by noting that if the
matter contained in an object scales as $L^d$, $L$ being its linear size,
its dimension will be $d$. For example, the area of a disk scales as $R^2$,
$R$ being its radius, indicating
it is a 2-dimensional object. The volume of a sphere scales as $R^3$,
implying it is a 3-dimensional object. For certain objects $d$ can be
noninteger, and is called its fractal dimension.}
\begin{figure}[h]
\subfigure[]{\includegraphics[width=5.9cm]{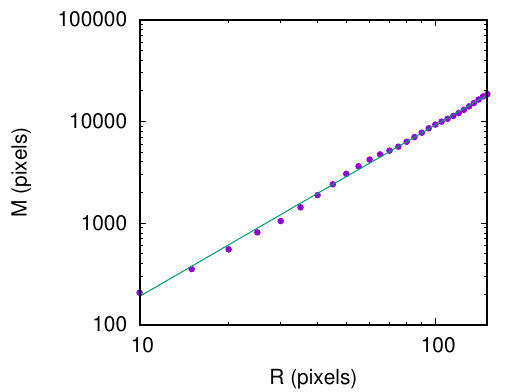}}
\subfigure[]{\includegraphics[width=5.9cm]{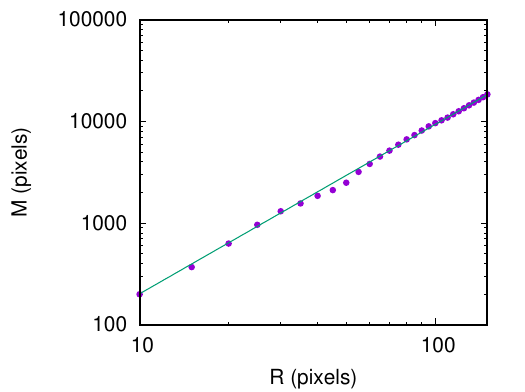}}
\subfigure[]{\includegraphics[width=5.9cm]{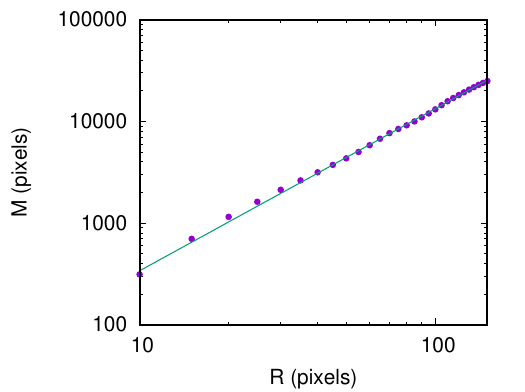}}
\caption{Dimension analysis of the patterns in Fig. \ref{antsdla1}.
The amount of chipped area inside a varying radius $R$ is caculated,
and plotted against $R$.
Analysis of \ref{antsdla1}(a): The best fit
line is a power law, matter {$M = (3.95\pm 0.35) R^{1.68\pm 0.02}$}. This indicates that
the pattern is a fractal with dimension 1.68.
Analysis of \ref{antsdla1}(b): The best fit line is a power law,
matter {$M = (4.38\pm 0.33)R^{1.66\pm 0.01}$}.
This implies that the pattern is a fractal with dimension 1.66.
Analysis of \ref{antsdla1}(c): the best fit line is a power law,
matter {$M = (8.71\pm 0.54)R^{1.59\pm 0.01)}$}, indicating that the pattern is a fractal
with dimension 1.59. {Goodness of fit is indicated by the asymptotic standard errors (ASE) mentioned with the fit values.}
}
\label{analysis1}
\end{figure}

A python code was used to analyze the patterns. Starting from a point
on the pattern, the ``mass'' of the white $M$ (contrasted with the reddish
paint), counted in terms of the number of 
pixels, is contained inside a circle of radius $R$ (in terms of number of pixels).
The center was manually identified in the patterns and the radius
was limited to remain inside the main body of the patterns.
{The program uses two reference colors: Pure red (255,0,0)
(in RGB language),
and pure white (255,255,255). The color of each pixel is read as (R,G,B).
Its color distances from pure red and pure white are calculated as
$d_{\text{red}}=\sqrt{(R-255)^2+G^2+B^2}$ and
$d_{\text{white}}=\sqrt{(R-255)^2+(G-255)^2+(B-255)^2}$.
Depending on which distance is shorter, the pixel is characterized as
either red or white.}
The amount of white stuff, {which is the number pixels characterized
as white,}
is then plotted against $R$ on a log-log scale to determine how the amount
of white stuff scales with $R$. If the amount of white stuff scales as
$M \sim R^d$, the object is considered to have a dimension $d$.

The pattern shown in Fig. \ref{antsdla1}(a) is the most spread out of the
patterns seen. The analysis in Fig. \ref{analysis1}(a) shows that it has
a dimension 1.68, indicating that the pattern is fractal in nature.
The pattern shown in Fig. \ref{antsdla1}(b) is somewhat directional. 
The analysis in Fig. \ref{analysis1}(b) shows that it has
a dimension 1.66, again indicating the fractal nature of the pattern.
The pattern in Fig. \ref{antsdla1}(c) is smaller in spread, but contains
more heavily eaten parts than others.
The analysis in Fig. \ref{analysis1}(c) shows that it has
a dimension 1.59, indicating that this pattern is also fractal in nature,
like the others.

\begin{figure}[h]
\centerline{\resizebox{10.0cm}{!}{\includegraphics{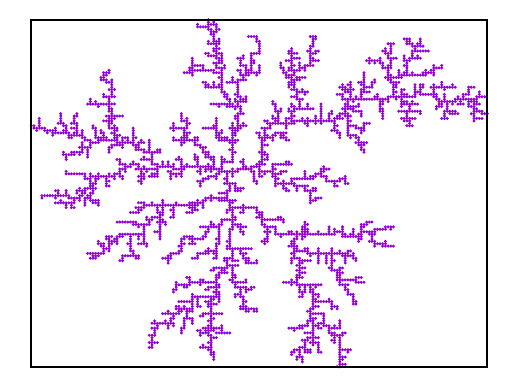}}}
\caption{The dynamics of the process was simulated by having a chip in
the paint in the center, and ants coming in from far away, until they
come across the chip, and bite off another bit from the existing chip.
The fractal dimension of this pattern comes out to be 1.71.  }
\label{dla}
\end{figure}

\section{Simulating ants dynamics}

{In order to verify that the collective activity of ants involving
biting off the paint on the wall was like a DLA, the process was simulated
on a square lattice with a single chipped site at the center. Random
walkers, representing foraging ants, were released sequentially
from random points on the boundary and allowed to perform unbiased
random walks. When a walker reached a site adjacent to the chipped
cluster, that site was added to the cluster, representing the removal
of a paint fragment by an ant. The walker was then removed, and a new
walker was released from the boundary. This procedure was repeated until
approximately 3000 chipping events had occurred.
The resulting pattern is shown in Fig. \ref{dla}. Its fractal dimension was
found to be 1.71, consistent with the known value for two-dimensional
DLA clusters \cite{mandelbrot}. This value is also close to the fractal
dimension of the most extended ant-generated pattern shown in 
Fig. \ref{antsdla1}(a). Another measure of the structural complexity of a
cluster is its interface width or ``roughness''. Assuming that the center
of the cluster is
at the origin of the coordinates, one divides the cluster into angular
bins. Average radius of the cluster is then defined as 
\begin{equation}
 \langle R\rangle = \frac{1}{n_{\theta}} \sum_{i=1}^{n_{\theta}} R(\theta_i),
\end{equation}
where $n_{\theta}$ is the number angular bins, and $R(\theta_i)$ the distance
of the farthest point in the $i$'th bin. Roughness is then defined as
\begin{equation}
 W = \sqrt{ \frac{1}{n_{\theta}} \sum_{i=1}^{n_{\theta}} (R(\theta_i)-\langle R\rangle)^2}.
\end{equation}
For the cluster depicted in  Fig. \ref{dla}, the above defined quantities
yield the values $\langle R\rangle = 55.08$ and $W = 20.05$.
Assuming that the roughness scales with the cluster size as $W \sim \langle R\rangle^{\alpha}$, the roughness exponent comes out to be $\alpha = 0.75$.
}

\begin{figure}[h]
\centerline{\resizebox{10.0cm}{!}{\includegraphics{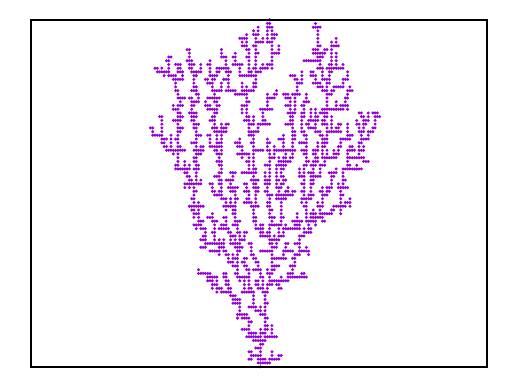}}}
\caption{A separate simulation done carried out to mimic the situation
depicted in Fig. \ref{antsdla1}(b), where the access of the chipped point in
the paint, to the ants, seems to be restricted. Here the random walkers
always started from above, and moved somewhat preferentially downwards.  }
\label{vdla}
\end{figure}

{To model the situation shown in Fig. \ref{antsdla1}(b), where ants
appear to
approach the chipped region only from above, the walkers were released
from a restricted region of the upper boundary. The resulting pattern,
shown in Fig. \ref{vdla}, qualitatively resembles the observed structure.
Its fractal dimension was found to be 1.65.}

\section{Conclusion}

This study examined the branching structures formed by ants biting paint
off a wall. A part of each structure was considered to be contained in a
circle with the center guessed from the pattern, and radius R. The amount of chipped paint
($M$) was counted in terms of the number of pixels. A python program was
used to determine dimension $d$ in the relationship $M \sim R^d$. The values of
$d$ for three different patterns were 1.68, 1.66, and 1.59, implying that
the patterns were fractal in nature.  This showed that the patterns
were governed by DLA.
{The evaluated dimensionality of the patterns varies probably
because the limited size of some patterns affects its accurate  estimation}.
In addition in many situations the ants 
are not able to approach the initial chip in the paint from all sides, hence
a deviation from the true DLA setting is to be expected. {Although DLA has been applied earlier to
things like microbial biofilm growth \cite{biofilm},} to our knowledge,
this study is the first to report
a case where DLA is observed in a phenomenon that involves {foraging}
live beings
guided by physical senses. A model for this phenonmenon was simulated,
and the results were in good agreement with the observations.

One might observe that the mechanism elucidated herein
diverges from the majority of DLA mechanisms, both theoretical
and empirical, in that no material undergoes aggregation in this
context. Rather, the material is progressively removed through incremental
erosion. {The way we understand how this process is equivalent
to the DLA mechanism is as follows. We assume that foraging ants behave
akin to a random walker. What is important in this part is that the
movement of the ants does not have any directional preference. In the 
DLA process when the random walker comes across any part of the aggregated
structure, it sticks there. Equivalently, when an ant comes across a
chipped paint region, it realizes that it need not search any further,
and that a paint piece can be bitten off right there. Thus the ant bites
off a piece and there is an addition to the eroded paint region. As more 
ants repeatedly participate in this process, there is an \emph{aggregation}
of the eroded paint region, exactly as there is aggregation of particles
in the conventional DLA process. These two processes, random walk, and 
immediate termination of the walker on encountering the aggregate,
are enough to generate
a DLA structure. We believe that is the reason why the
structures created here are so similar to DLA patterns. 
Anything that would affect the random motion of ants close to the
region of chipped paint, may affect the observed patterns. For example,
if there are pheromone trails inside the structure of the chipped paint
region, the ants may repeatedly go to the same parts. This may lead to
deviation in the observed patterns from standard DLA patterns. From the
observed patterns it appears that the pheromone trails only guide the
ants to the broad region, but do not play significant role inside the
generated eroded paint structure.}

We found
it very interesting that a phenomenon involving live beings gives a result
very similar to a phenomenon based on physical particles governed by
simple physical laws.  It would be interesting to see if DLA like
patterns exist in other situations involving other animals.

\section*{Acknowledgement}
The authors thank Saumitra Chatterjee for bringing the phenonmenon to
our attention.

\section*{Funding information}
No funding was used in this work.

\section*{Data availability}
The authors declare that the data supporting the findings of this
study are available within the paper, and no additional data was
generated in this work.\\
A video showing the ants at work, and the
python program used to analyze the images are available at the
repository \href{https://doi.org/10.5281/zenodo.23217831}{10.5281/zenodo.23217831}.

\section*{Conflict of interest}
The authors have no conflict of interest to declare.

\end{document}